\documentclass[letterpaper]{article} 
\usepackage{aaai24}  
\usepackage{times}  
\usepackage{helvet}  
\usepackage{courier}  
\usepackage[hyphens]{url}  
\usepackage{graphicx} 
\usepackage{natbib}  
\usepackage{caption} 
\usepackage{algorithm}
\usepackage{algorithmic}
\usepackage{svg}

\usepackage{newfloat}
\usepackage{listings}
\DeclareCaptionStyle{ruled}{labelfont=normalfont,labelsep=colon,strut=off} 
\floatstyle{ruled}
\newfloat{listing}{tb}{lst}{}
\floatname{listing}{Listing}
\usepackage{adjustbox}

\title{Predicting ATP binding sites in protein sequences using Deep Learning and Natural Language Processing}
\author {
    Shreyas V\textsuperscript{\rm 1,2},
    Swati Agarwal\textsuperscript{\rm 1}\thanks{Corresponding Author},
}
\affiliations {
    \textsuperscript{\rm 1}Department of Computer Science and Information Systems\\
    \textsuperscript{\rm 2}Department of Chemistry\\
    BITS Pilani, Goa Campus\\
    \{f20201350,swatia\}@goa.bits-pilani.ac.in
}

\begin{document}

\maketitle

\begin{abstract}
Predicting ATP-Protein Binding sites in genes is of great significance in the field of Biology and Medicine. The majority of research in this field has been conducted through time- and resource-intensive 'wet experiments' in laboratories. Over the years, researchers have been investigating computational methods computational methods to accomplish the same goals, utilising the strength of advanced Deep Learning and NLP algorithms. In this paper, we propose to develop methods to classify ATP-Protein binding sites. We conducted various experiments mainly using PSSMs and several word embeddings as features. We used 2D CNNs and LightGBM classifiers as our chief Deep Learning Algorithms. The MP3Vec and BERT models have also been subjected to testing in our study. The outcomes of our experiments demonstrated improvement over the state-of-the-art benchmarks.
\end{abstract}

\section{Introduction}

Adenosine Triphosphate (ATP) is a ubiquitous organic molecule present in all known life forms, ranging from rudimentary bacteria to the human species. ATP plays a crucial role in various essential biochemical processes within living cells, including intracellular signalling, DNA/RNA synthesis, and protein transport, as evident by studies conducted by \citet{novak2003atp}, \citet{enomoto1981atp} and \citet{ruprecht2019molecular} respectively. 
The phrase "molecular unit of currency" is commonly used to describe its role in facilitating intra-cellular energy transfer. According to \citet{narunsky2020evolution}, ATP molecules engage in interactions with a diverse range of proteins, thereby facilitating the discharge of requisite chemical energy for optimal protein functionality \cite{alberts2002}. The examination of these interactions and precise forecasting of ATP binding sites within a particular sequence is informative for the annotation of protein function and the advancement of pharmaceuticals \cite{Schmidtke2010-qy, Sirimulla2013-ol, Verdonk2004-ig, Amari2006-yt}. The significance of protein-ligand interactions cannot be overstated in various biological processes, including but not limited to DNA replication and transcription, membrane transportation, and cellular respiration \cite{Verteramo2019, doi.org/10.1111/cbdd.13648}. Precisely determining the locations of binding sites within proteins is valuable for annotating protein function and developing new drugs to treat various ailments such as cancer \cite{Yuan2018-bn}, diabetes \cite{Miller2020-dl}, and Alzheimer's disease \cite{Sun2019-ys}. 
The ligand molecule known as ATP plays a crucial role in cell biology by serving as both an energy source and a coenzyme \cite{Maxwell2003-hu}. Proteins engage in interactions with one another by means of protein-ATP binding residues present in protein sequences. This interaction results in the provision of chemical energy to proteins through hydrolysis, which can be utilised for a multitude of protein functions \cite{YU2013180, zhang2012predicting}.

Considerable experimental work has been conducted in wet-lab settings to ascertain the precise sites of protein-ATP binding residues, utilising techniques such as X-ray crystallography \cite{doi:10.1126/science.1217737} and Nuclear Magnetic Resonance (NMR) \cite{doi:10.1073/pnas.0610313104}. The wet-lab experiments are frequently limited in their application to the postgenomic era's large-scale protein sequences due to their cost-intensive and time-consuming nature \cite{10.1093/bioinformatics/bty816}. The ordered sequences of proteins render them amenable to effective application of NLP techniques.

In light of these conditions, the utilisation of computational methodologies for forecasting protein-ATP binding residues is has gained increasing interest among scholars, owing to the advancements in artificial intelligence and machine learning. The computational prediction methods can be classified into two categories based on the protein features involved. The first category is sequence-based methods, which utilise protein sequence information to derive features. The second category is structure-based methods, which derive features from structural protein information. As of November 4, 2020, it was observed that the quantity of protein structures present in the Protein Data Bank \cite{10.1093/nar/28.1.235} was comparatively lesser, standing at approximately 170,594, in contrast to the Swiss-Prot database \cite{10.1093/nar/24.1.21}, having around 563,552 structures. This difference in numbers can be attributed to the fact that the detection of 3-dimensional protein structures is a more challenging task as compared to the identification of protein sequence information. Hence, the utilisation of sequence information for the prediction of protein-ATP binding residues holds significant potential for broader applications.

The literature indicates that conventional wet-lab methods exhibit a notable degree of efficacy and precision \cite{cala2014nmr}. Nevertheless, these experiments are also economically unfeasible and require a significant amount of time. The rapid rate at which new proteins are being discovered necessitates the development of effective computational techniques for discerning the inherent patterns within their sequences and forecasting their bindings. Until recently, the majority of approaches utilised the structural information of both the protein and ligand. Furthermore, various machine learning methodologies have been utilised to analyse sequence features such as secondary structure and solvent accessibility, as demonstrated in previous studies \cite{hu2018atpbind,zhang2012predicting}. \citet{chauhan2009identification} utilised SVM classifier and developed ATPint, a predictor of ATP based on sequences by incorporating Position Specific Scoring Matrices (PSSMs). In addition to the PSSMS, \citet{chen2011atpsite} integrated the forecasted secondary structure, predicted dihedral angles, and relative solvent accessibility. \citet{yu2013targetatpsite} introduced a computational tool called TargetATPsite. This tool employs a classifier ensemble and generates sparse representations of the PSSM profiles. \citet{hu2018atpbind} suggested the ATPbind model which integrated the results of two predictors based on templates and sequence characteristics.

In recent times, there has been a remarkable demonstration of the effectiveness of deep learning techniques in sequence modelling across diverse domains, including Computer Vision \cite{voulodimos2018deep}, Natural Language Processing \cite{young2018recent}, and Bioinformatics \cite{min2017deep}. \cite{kusuma2019prediction} and \cite{song2020novel}, utilised deep learning techniques, specifically 2D Convolutional Neural Networks (CNNs), to predict ATP-binding sites based on protein sequence profiles generated by PSSMs. The authors utilised PSSM profiles as the main feature vectors and constructed two classification networks, namely a residual-inception-based predictor and a multi-inception-based predictor. 

The prevailing trend in ATP-binding prediction research involves the utilisation of distinct features and the application of prediction models as opaque entities. A significant hurdle in the post-genomic epoch is the provision of functional annotations for a vast quantity of proteins that result from genome sequencing initiatives. The interaction of numerous proteins with small molecules or ligands is pivotal for their functionality. ATP is a significant ligand that serves as a crucial coenzyme in the operation of numerous proteins. It is imperative to devise a methodology for discerning ATP-interacting residues within ATP binding proteins (ABPs) to gain insight into the mechanism of protein-ligand interactions.

In contrast to the existing literature, our paper presents an efficient and effecting model, while also being easily interpreted. This model utilises both sequence-based information, specifically PSSMs, and secondary structure information derived from proteins. NLP techniques are employed to represent protein sequences as n-grams, which are subsequently utilised as features. 

\section{Experimental Datasets}\label{sec:dataset}
\begin{figure}
    \centering
    \includegraphics[width=\linewidth]{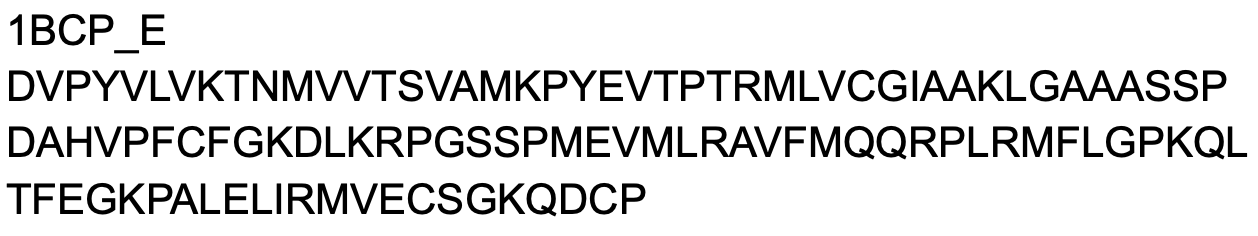}
    \caption{A Sample Snapshot of a Protein Sequence Present in our Dataset.}
    \label{fig:Sequence}
\end{figure}

\begin{figure}
    \centering
    \includegraphics[width=\linewidth]{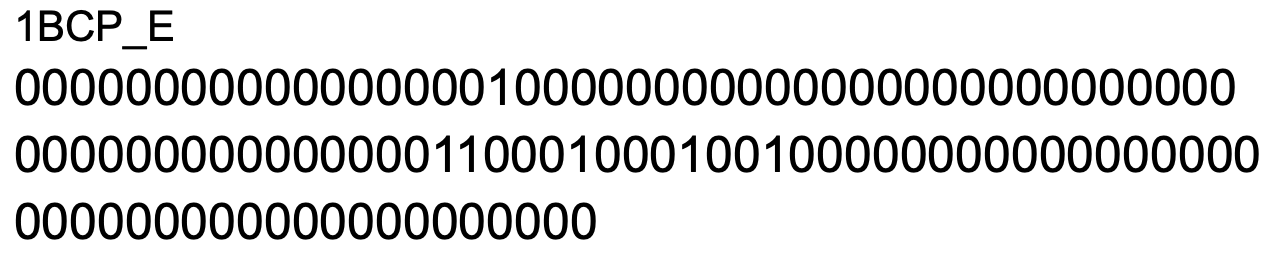}
    \caption{A Snapshot of a Binary Encoded Labels Denoting the Presence and Absence of Protein Binding Sites.}
    \label{fig:Labels}
\end{figure}

In order to conduct our experiments and validate the proposed features, we acquired three sets of open-source datasets from a public repository and the current state-of-the-art literature. The ATP-168, ATP-227, and PATP-388 datasets were chosen for experimentation due to their broad use in the research community and availability for benchmarking and comparison purposes. The datasets are supplied as flat files, with each file containing a fixed number of protein sequences. Each dataset file contains three fields: Protein Sequence ID, Protein Sequence of Amino Acids, and Binary Encoded Labels. The binary encoded labels indicate the presence or absence of protein binding at a specific amino acid, where 1 denotes binding and 0 indicates the absence of binding. A large number of protein files make up the dataset. Each file contains amino acid sequences and their related labels. As illustrated in Figure \ref{fig:Sequence}, the amino acid sequence 1BCP\_E is made up of a series of amino acids, with each letter representing a separate unit. Figure \ref{fig:Labels} shows the labels allocated to each amino acid. The value of 0 indicates the absence of protein binding at the respective site, while the value of 1 signifies the presence of protein binding at that specific site.

\subsubsection{PATP 388 and PATP-41} \label{sec:patp_388}
In 2018, \citet{hu2018atpbind} extracted 2144 ATP binding proteins from PDB database. These binding protein had target annotations. \citet{hu2018atpbind} removed the redundant sequences resulting in 429 unique sequences. These sequences were split into two sets: 388 and 41, for training and testing respectively. PATP-388 contains 388 protein sequences while PATP-41(TEST) includes 41 protein chains. More specifically, PATP-388 ccomprises 5657 ATP binding residues (i.e., positive samples) and 142086 non-ATP binding residues (i.e., negative samples). We use the same dataset split for our experiments, i.e., 388 sequences for training and 41 sequences for testing purpose. 

\subsubsection{ATP-227 and ATP-17} \label{sec:atp_277}
In 2011, \citet{chen2011atpsite} created the ATP-227 dataset containing 227 protein chains. The binding residue is defined if at least one of its non-hydrogen atoms is less than 3.9 Å away from a non-hydrogen atom of the ATP molecule. As a byproduct, authors created the ATP-17 dataset consisting of ATP-binding protein chains released after March 10, 2010. To avoid biases in the testing dataset, they reduce the maximal pairwise sequence identity in ATP-17 to 40\%. Thus, if a given chain shares $>40\%$ identity with a chain in ATP-227, then the chain from ATP-17 was removed. This process assures that ATP-17 is independent of ATP-227 and can be used as a testing set for models that are trained on ATP-227. Consequently, 17 ATP-binding protein chains remain in the ATP-17 testing set. 

\subsubsection{ATP-168} \label{sec:atp168}
This dataset was originally used by \citet{chauhan2009identification} which extracted 360 ATP-binding protein chains from the SuperSite database \cite{bauer2009supersite}. To eliminate the duplicate biases, redundant sequences with a pairwise sequence identity of more than 40\% were deleted. Following that, the proteins were evaluated on Ligand Protein Contact software \cite{sobolev1999automated} to ensure their validity as an ATP-binding protein. The protein disqualified by the software were removed, resulting in a final dataset of 168 non-redundant proteins. 

We divide our training datasets (PATP-388, ATP-227, and ATP-168) into training and validation by randomly sampling 10\% of the dataset for use as validation sets. The remaining 90\% of the dataset was utilised for training purposes. Evident from the data presented in Table \ref{table:datasetstatstics}, the datasets exhibit a significant degree of skewness or imbalance, with a significant disparity between the number of Negative Samples and Positive Samples.


\begin{table}[!htbp]
    \centering
    \resizebox{\linewidth}{!}{
        \begin{tabular}{|l|l|l|l|l|l|}
        \hline
        \multicolumn{3}{|c|}{} & \multicolumn{2}{|c|}{Samples} & \\\hline
        Dataset & Type & No. of seq & Positive & Negative & Ratio \\ \hline
        ATP-168 & Training & 168 & 59225 & 3104 & 19.08\\\hline
        ATP-227 & Training & 227 & 3393 & 80409 & 23.70 \\ \hline
        ATP-17 & Testing & 17 & 248 & 6974 & 28.12 \\ \hline
        PATP-388 & Training & 388 & 5657 & 142086 & 25.12 \\ \hline
        PATP-41(TEST) & Testing & 41 & 674 & 14159 & 21.01 \\ \hline
        \end{tabular}
    }
    \caption{Statistical decomposition of the datasets}
    \label{table:datasetstatstics}
\end{table}
\section{Proposed Methodology}
The proposed method is a multi-step process which consists of feature engineering, addressing the data imbalance, and classification model. We validate our proposed approach by using combinations of several features' matrix and classification models and compare them against state-of-the-art techniques. We discuss each of these phases in the following subsections:
\subsection{Feature Engineering}\label{subsec:features}
The influence of nearby residues on the behaviour and characteristics of protein residues was investigated using a sliding window technique. A sliding window of size W contains the properties of the target residue as well as those of the $\frac{L-1}{2}$ residues to its left and right. We discuss these features in the following subsections:
\subsubsection{Position Specific Scoring Matrix (PSSM)}
PSSMs are used to provide evolutionary conservation data about a specific protein sequence. Creating replacement scores entails providing a numerical value to each position in a multiple sequence alignment. A positive score shows that there has been an increase in the frequency of amino acid substitution over what would be predicted by chance. A negative score, on the other hand, implies that the substitution occurred with less frequency than expected. Table \ref{tab:sample_pssm_profile} shows an example of a PSSM profile. Prior studies have proved the significance of PSSMs application in predicting ligand binding sites \cite{cala2014nmr} and structure prediction, as demonstrated by \citet{uhl2019graphprot2}. The PSSMs are created by running the PSI-BLAST algorithm three times against the UniProt database, as reported by \citet{altschul1997gapped}. The matrix is L x 20, where L denotes the length of the protein sequence. The number 20 represents the total number of different amino acids known to exist in the dataset. Through the use of a modified sigmoid function, the matrix values were normalised to adhere to the interval [-1, 1]. For every given value x within the PSSMs:
\[ x_{norm}  =  \frac{2}{(1 + e^{(-x/2)})}-1\]
\begin{table}[ht]
    \centering
    \resizebox{\linewidth}{!}{
    \begin{tabular}{|c|c|c|c|c|c|c|c|c|c|c|c|c|c|c|c|c|c|c|c|c|c|}
    \hline
    & & \multicolumn{20}{|c|}{Amino Acid}\\
    
    \hline
  & & A & R & N & D & C & Q & E & G & H & I & L & K & M & F & P & S & T & W & Y & V \\
         \hline
         1& G & 0 & -2 & 0 &-1&-2&-2&-2& 6&-2&-4&-4& -2&-3& -3&-2& 0&-2& -2& -3&-3\\
         2& S & 1 & -1 & 1 & 0&1& 0& 0& 0&-1&-2&-2& 0&-1& -2& -1& 4& 1&-3&-2&-2 \\
         3& R & -1& 5 & 0 &-2&-3&1& 0&-2& 0&-3&-2& 2&-1&-3&-2&-1&-1&-3& -2& -2 \\
         4& E & -1& 0& 0& 1& 4& 2& 5&-2& 0&-3&-3& 1&-2&-3&-1& 0&-1&-3&-2&-2 \\
         5& F & -2&-3&-3& -3& -2&-3&-3&-3&-1& 0& 0&-3&0 &6&-4& -2&-2& 1& 3&-1 \\
         6& D & -1&-1& 5& 4&-3&-1& 0&-1&-1&-4& -4& -1&-3& -4&-2& 2& 0&-4&-3&-3\\
         7& Q & -2& 5&-1&-2& 1& 2& 0&-3&-1&-3&-3& 3&-2&-3&-2&-1&-2& 5&-1&-3 \\
         8& K & -2& 5&-1& -2&-4& 1& 0&-3&-1&-3& -3& 4&-2&-4& 0&-1& 0&-3&-2&-3 \\
         9& I & -1&-3&-4&-4&-2&-3&-3&-4&-4& 3& 3&-3& 1&-1&-3&-3& -1&-3& -2& 3 \\
         10& G& -1& -3& -1& -2& -3& -3& -3& 7& -3& -5& -5& -2& -4& -4& -3& -1& -2& -3& -4& -4 \\
         \hline
    \end{tabular}}
    \caption{Sample PSSM Profile, the rows represent some of the different protein sequences that were sampled}
    \label{tab:sample_pssm_profile}
\end{table}
\subsubsection{FastText vectors}
FastText library was developed by Facebook for word representation learning and text classification \cite{bojanowski2017enriching}. It can be used to train several language models, such as skip-gram and CBOW, with the desired sampling technique, loss function, and hyperparameters. Various studies have applied Word2Vec technique \cite{mikolov2013efficient} to construct embeddings for biological and medical data \cite{wu2019ptpd,wang2018comparison}. More recently, \citet{le2019identifying,le2019classifying} have successfully employed FastText to express biological sequences. FastText differs from the Word2Vec approach in that it considers subwords in addition to words, allowing for training on smaller datasets and generalisation to previously unseen words. We represent protein sequences as continuous overlapping n-grams. For example, when n = 3, the sequence “KSLMFFI” would be represented as the “sentence” $<$ KSL, SLM, LMF, MFF, FFI $>$ . Each protein in the UniProt database is represented as a sentence consisting of overlapping n-gram “words”, and a FastText model was trained using this dataset. The trained model was then used to generate 100-dimensional phrase vectors for each window of the input protein sequence. To extract the most subword information, the n parameter was set to 5, and the $min_n$ and $max_n$ parameters, which are the minimum and maximum lengths of character n-grams, were set to 1 and 5, respectively. The sentence vectors generated by this trained model were used as input for the pipeline's following stage.

\subsubsection{Predicted Secondary Structure}
Previous research on this subject has shown that incorporating the anticipated secondary structure of the protein sequences improves the model's performance. This could be because the protein's ATP-binding residues assume a certain secondary and tertiary structure in order to bind to their ligands and fulfil their biological tasks. PSIPRED is a programme that predicts the secondary structure of protein sequences \cite{mcguffin2000psipred}, producing a set of probabilities that the residue is part of a helix, sheet, or coil. The secondary structural feature's dimensions are thus W3, where W is the size of the sliding window. Table \ref{tab:sample_psipred_output} shows an example of PSIPRED output.
\begin{table}[!htbp]
    \centering
    \begin{tabular}{c c c c c c}
    \textbf{S.No} & \textbf{Protein} & \textbf{C/H/S} & \textbf{C} & \textbf{H} & \textbf{S} \\
    \hline
        1 & T & C & 0.999 & 0.001 & 0.001 \\
        2 & G & C & 0.953 & 0.013 & 0.034 \\
        3 & V & C & 0.742 & 0.080 & 0.191 \\
        4 & K & C & 0.735 & 0.147 & 0.162 \\
        5 & I & H & 0.647 & 0.938 & 0.012 \\
        6 & R & H & 0.073 & 0.919 & 0.020 \\
        7 & D & H & 0.837 & 0.962 & 0.009 \\
        8 & L & H & 0.864 & 0.936 & 0.008 \\
        9 & V & H & 0.875 & 0.921 & 0.014 \\
        10 & K & H & 0.182 & 0.803 & 0.021 \\
        11 & H & H & 0.451 & 0.542 & 0.019 \\
    \end{tabular}
    \caption{Sample PSIPRED Output. Helix(H), Sheet(S), Coil(C)}
    \label{tab:sample_psipred_output}
\end{table}
\subsection{Addressing Data Imbalance}
The three datasets have a large imbalance, with negative (non-ATP-binding residues) instances significantly outnumbering positive samples (ATP-binding residues). This issue may cause the model to regularly forecast the negative class without making any substantial inferences. To solve the issue of imbalanced datasets, various techniques can be used. Oversampling is a strategy that includes reproducing instances from an underrepresented class, which might result in over-fitting in some scenarios. Under-sampling is the removal of samples from the majority class, which can result in the loss of important information.

\subsubsection{SMOTE algorithm:} In contrast to the existing studies on predicting ATP-binding, we use Synthetic Minority Oversampling Technique (SMOTE) to address the data imbalance problem \cite{Chawla_2002}. Rather than repeating the minority class instances, the SMOTE algorithm keeps all training information and generates synthetic samples to balance the dataset, as opposed to under-sampling. This method successfully reduces the risk of the classifier over-fitting. The ATP-388 dataset had a negative-to-positive sample ratio of 25.12 prior to the installation of SMOTE. This ratio was reduced to 18.965 once SMOTE was applied.
\citet{song2020novel} has shown ablation studies on different methods of addressing data imbalances. And their results revealed that SMOTE resulted in improvement of the MCC from 0.434 to 0.480 on the ATP-168 dataset and from 0.476 to 0.535 on ATP-227 Dataset. Thus we acknowledge that using smote is a proven may way of dealing with imbalances in the datasets used in this study

\subsubsection{LightGBM:} 
The LightGradient Boosting Decision Tree is an iterative decision tree-based technique that may be used for classification and regression \cite{lightgbmpaper}. Given a training dataset, a negative gradient of the loss function from the model output is obtained after each gradient enhancement step. In the decision tree, the feature with the greatest information gain is then chosen to partition each node. 
While previous literature used 2D CNNs categorization\cite{kusuma2019prediction}; we propose to use PSSMs + PSIPRED + FastText vectors as features. The ReLU activation is used by every trainable layer. Throughout the model, dropout layers are employed to prevent overfitting and as an implicit ensembling approach. \citet{song2021prediction} shows that LightGBM may be used in conjunction with a CNN Classifier to obtain state-of-the-art results for predicting ATP-Protein Binding.

\subsubsection{BERT:} Bidirectional Encoder Representations from Transformers (BERT) understand the context of a word by looking at its surroundings in both directions, which helps the model generate more nuanced language interpretations \cite{devlin2019bert}. BigBird is a sparse-attention based transformer that extends BERT and other Transformer-based models over much longer sequences \cite{big_bird_paper}. Furthermore, BigBird includes a theoretical understanding of the capabilities of a full transformer that the sparse model can handle. 
We used this approach to address the significant length of protein sequences in the PATP-388 dataset, which ranged from 65 to 3005 proteins.

\subsubsection{MP3Vec:} The Multi Purpose Protein Prediction Vector (MP3Vec) representation is built from a protein's sequence and PSSM profile. It is built by using all available high resolution protein structural data to aid in 'transfer learning'. The transfer learning in MP3Vec overcomes the 'small data' problem in developing predictive models for biomacromolecular interactions \cite{mp3vec}. We used MP3Vec to produce the feature vector for a given protein sequence, which was then used to train our proposed MP3Vec-Based model.

\subsection{Proposed Model}\label{sec:model}
\begin{figure}[!htbp]
    \centering
    \includegraphics[width=1\linewidth]{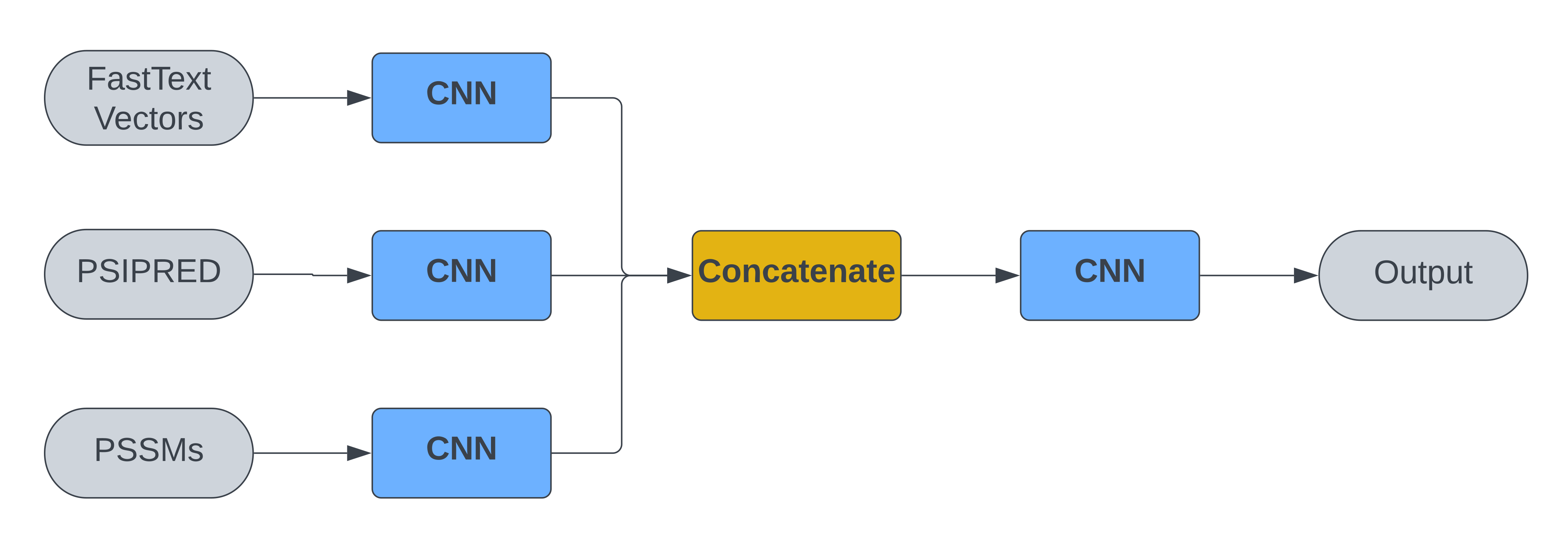}
    \caption{Proposed Model Architecture}
    \label{fig:model_architecture}
\end{figure}

A deep neural network is used to forecast a residue's ATP-binding status. Three independent features are processed by three separate branches, and the resulting representations are merged to generate the final output. The PSSM and PSIPRED features both make use of 1D convolutional layers. These can be used to exploit one-dimensional inputs for their spatial proximity, similar to how 2D convolutional layers work, leading in the extraction of important representations from brief data sequences. For the goal of simulating sequential data, LSTM and GRU units have traditionally been used. According to research, 1D CNNs perform similarly to the aforementioned models in circumstances where the sequence is not overly long, while also being substantially more expeditious and computationally efficient. Due to the lack of spatial information that can be used, the FastText sentence vector features are frequently combined with a Dense network. The structure of the model is depicted in Figure \ref{fig:model_architecture}. A more comprehensive model is explicated in Figure \ref{fig:detailed_model_architecture} under appendix.
The ReLU activation is used by every trainable layer. Throughout the model, dropout layers are employed to prevent overfitting and as an implicit ensembling approach. The Adam optimizer is used to minimise the binary cross-entropy loss as an objective function.
\section{Performance Evaluation}
We assess the proposed method's overall performance using four commonly used evaluation criteria: overall accuracy (ACC), sensitivity (SEN), specificity (SPE), and Matthews correlation coefficient (MCC). These standards for evaluation are frequently used in BioInformatics research \cite{biology9100325_2020, ijms21239070_2020} to reveal classification performance. MCC is beneficial for imbalanced datasets with large class differences. It evaluates the model's performance based on sensitivity (recall) and specificity for all four classification methods (TP, TN, FP, FN). The MCC ranges from -1 to +1. +1 means flawless prediction, 0 means random prediction, and -1 means total discrepancy between projected and actual classes.

{\scriptsize
\[MCC = \frac{TP*TN-FP*FN}{\sqrt{(TP + FP)(TP + FN)(TN + FP)(TN + FN)}}\]}
The evaluation criteria are threshold-dependent and hence indicate predicted performance within a specific threshold. To establish a fair comparison between our suggested methodology and other sequence-based prediction methods, we used the same evaluation criteria as the current literature. \cite{yu2013targetatpsite, YU2013180, hu2018atpbind, HU2016363, 10.1093/bioinformatics/btr657}.
\section{Experimental Results}
\subsection{Effect of window size}
\begin{figure}[t]
    \centering
    \includegraphics[width=1\columnwidth]{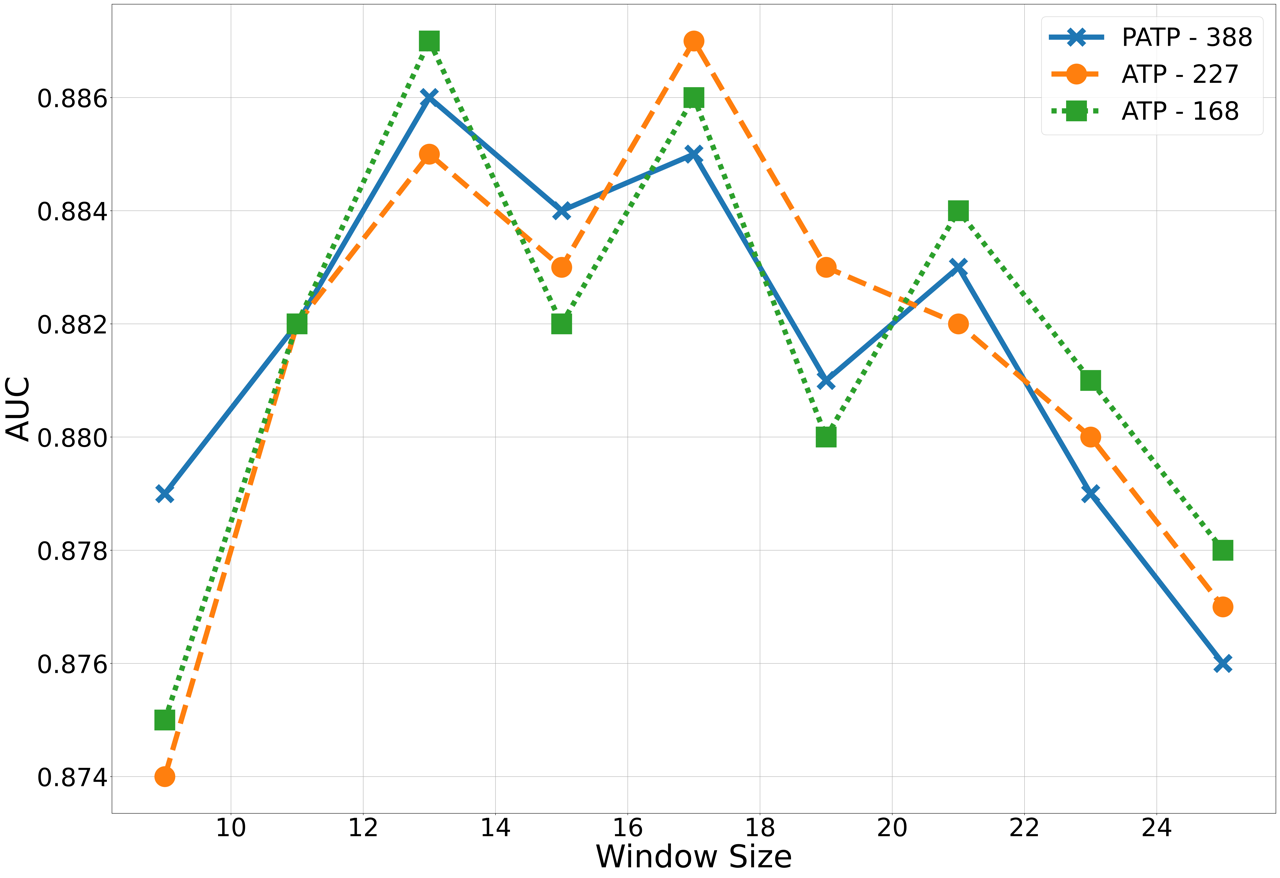}
    \caption{Effect of window size on AUC}\label{fig:auc_windows}
  \end{figure}

To find an acceptable window size, we trained models for a range of sizes from 9 to 25 utilising solely the PSSM capabilities of the datasets. A 90-10 train-validation split is used to understand the impact of window size on the performance of the model. Figure \ref{fig:auc_windows} shows the variation in the Area Under ROC Curve for all three datasets across various window sizes. Based on the results, we find that all three datasets share the similar pattern (but different absolute values) for AUC. However, the highest AUC is achieved for the window size 17. Thus we use W=17 for all further experiments to capture sufficient context around the target residue without comprising with the computational cost. It is interesting to see that since ATP-168 and ATP-227 being derived from PATP-388, the datasets share the same pattern for varying window sizes. But the performance value vary due to different number and sequences of amino acids in the data. 
\subsection{Ablation studies on features}
We perform an ablation study on the features and evaluate the performance of the model against various combinations of the features to identify their importance in the prediction. For all three datasets and different combinations of features, a five-fold cross-validation was performed and the AUC is reported in Table \ref{tab:auc_features}.
\begin{table}[!htbp]
    \centering
    \small
        \begin{tabular}{|l|c|c|c|}
            \hline
            Dataset/Features & ATP-168 & ATP-227 & PATP-388\\
            \hline
            PSSM & 0.871 & 0.882 & 0.886\\
            PSSM + SS & 0.879 & 0.889 & 0.892\\
            PSSM + FastText & 0.873 & 0.883 & 0.888\\
            PSSM + SS + FastText & \textbf{0.882} & \textbf{0.893} & \textbf{0.899}\\
            MP3Vec-Based & 0.880 & 0.888 & 0.889 \\
            BERT-Based & 0.851 & 0.878 & 0.887 \\\hline
        \end{tabular}
    \caption{The AUC scores of Proposed Model with different combinations of features}
    \label{tab:auc_features}
  \end{table}
\begin{table*}[!htbp]
\centering
\small
\begin{tabular}{|l|l|c|c|c|c|c|}
    \hline
    Data & Method & Accuracy & Sensitivity & Specificity & MCC & AUC\\\hline
    ATP-17 & ATPint \cite{chauhan2009identification} & 0.665 & 0.512 & 0.660 & 0.066 & 0.606\\
    ATP-17 & ATPsite \cite{chen2011atpsite} & 0.969 & 0.367 & 0.991 & 0.451 & 0.868\\
    ATP-17 & NsitePred \cite{ijms21239070_2020} & 0.967 & 0.460 & 0.985 & 0.476 & 0.875 \\
    ATP-17 & TargetATPsite \cite{yu2013targetatpsite} & 0.972 & 0.458 & 0.991 & 0.530 & 0.882\\
    ATP-17 & TargetNUCs \cite{biology9100325_2020} & 0.975 & 0.516 & 0.992 & 0.584 & – \\
    ATP-17 & Song et al. \cite{song2020novel} & 0.975 & 0.549 & \textbf{0.993} & 0.595 & 0.922\\
    ATP-17 & Song et al. \cite{song2021prediction} & \textbf{0.978} & \textbf{0.589} & 0.992 & \textbf{0.639} & \textbf{0.925}\\ 
    \hline
    ATP-17 & PSSM + SS + FastText  & 0.976 & 0.532 & 0.981 & 0.547 & 0.913\\
    ATP-17 & MP3Vec-Based & 0.970 & 0.550 & 0.985 & 0.563 & 0.915 \\
    ATP-17 & BERT-Based & 0.951 & 0.522 & 0.980 & 0.539 & 0.902 \\
    
    \hline
    PATP-41(TEST) & NsitePred \cite{ijms21239070_2020} & 0.954 & 0.467 & 0.977 & 0.456 & 0.852\\
    PATP-41(TEST) & TargetATPsite \cite{yu2013targetatpsite} & 0.968 & 0.413 & 0.995 & 0.559 & 0.853\\
    PATP-41(TEST) & TargetNUCs \cite{biology9100325_2020} & 0.972 & 0.469 & 0.997 & 0.627 & 0.856\\
    PATP-41(TEST) & ATPseq \cite{hu2018atpbind} & 0.972 & 0.545 & 0.993 & 0.639 & 0.878\\
    PATP-41(TEST) & Song et al. \cite{song2020novel} & 0.972 & 0.494 & 0.995 & 0.626 & 0.896\\
    PATP-41(TEST) & Song et al. \cite{song2021prediction} & 0.973 & \textbf{0.497} & \textbf{0.996} & \textbf{0.642} & 0.902 \\ 
    \hline
    PATP-41(TEST) & PSSM + SS + FastText & \textbf{0.981} & 0.476 & 0.971 & 0.532 & \textbf{0.911}\\
    PATP-41(TEST) & MP3Vec-Based & \textbf{0.981} & 0.468 & 0.986 & 0.555 & 0.903\\
    PATP-41(TEST) & BERT-Based & 0.961 & 0.455 & 0.992 & 0.543 & 0.898\\
    \hline
\end{tabular}
\caption{Evaluation performance of the proposed models tested on ATP-17 and PATP-41(TEST) datasets. The table also illustrates the comparison with existing and benchmarking methods from the previous studies}
\label{tab:performance_atp17}
\end{table*}
The results reveal that combining all three characteristics produces the best results, and that the secondary structure predictions given by PSIPRED are more significant than the FastText vectors. It has been discovered that using MP3Vec produces results that are similar to the model created by including all three characteristics. The model based on BERT exhibits a slightly inferior performance in ATP-168, albeit approaching the level of performance of the model constructed utilising all three features. The potential cause of this occurrence could be attributed to the requisite of substantial resources for a BERT-like model to produce adequate results.
\subsection{Model Performance and Benchmarking}
\begin{figure}[!htbp]
    \centering
    \includegraphics[width=1\linewidth]{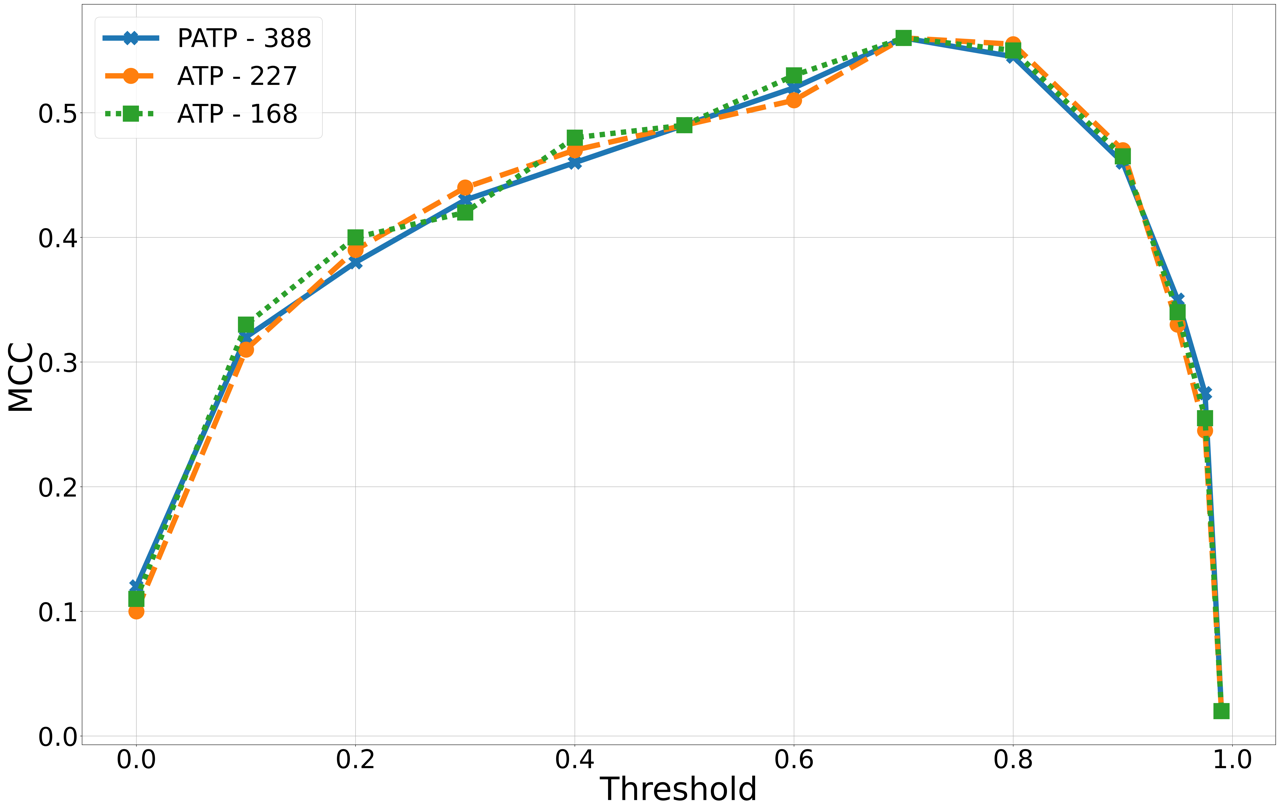}
    \caption{Threshold vs. MCC for PATP-388,  ATP-227 \& ATP-168}
    \label{fig:thres_mcc}
\end{figure}

It is important to acknowledge that the metrics in question may lack significance when considered in isolation, owing to the disproportionate distribution of data. The results of predicting an ATP-binding site can vary significantly depending on the chosen cutoff threshold. The threshold is selected to optimise the Matthews correlation coefficient (MCC) on the validation dataset, and subsequently applied to compute the remaining performance metrics. Figure \ref{fig:thres_mcc} shows that for all three datasets, the MCC score is maximized at a threshold of approximately 0.7 which is further used to calculate the other metrics.
Table \ref{tab:performance_atp17} shows the performance results of the proposed models and the existing literature on ATP-17 and PATP-41 test datasets. The results indicate that the our model exhibits superior accuracy and area under curve metrics compared to previously suggested models when evaluated on the PATP-41(TEST) dataset. The findings indicate that the MP3Vec-Based model exhibits a comparable performance to the proposed method. The observed phenomenon could potentially be attributed to the similar processing methodology employed by MP3Vec during the generation of its characteristic vectors. The performance of the BERT-Based model is comparable; however, it falls short in matching the performance of both the proposed method and the MP3Vec based model. We observe a similar pattern in ATP-17 dataset. \citet{song2021prediction} has better performance on ATP-17 dataset compared to this study due to the fact that \citet{song2021prediction} was tuned for ATP-227 and ATP-168 whereas this study is tuned for the PATP-388 dataset. However, while tested on ATP-168, ATP-227, and ATP-17- we obtained comparable results. Note that these datasets are not only different in sizes but also the imbalances of binding sites as demonstrated in Figures \ref{fig:residue_freq}  and \ref{fig:residue_freq_test}. 

\subsection{Analysis of ATP-binding residues}
\begin{figure}[!htbp]
    \centering
        \includegraphics[width=1\linewidth]{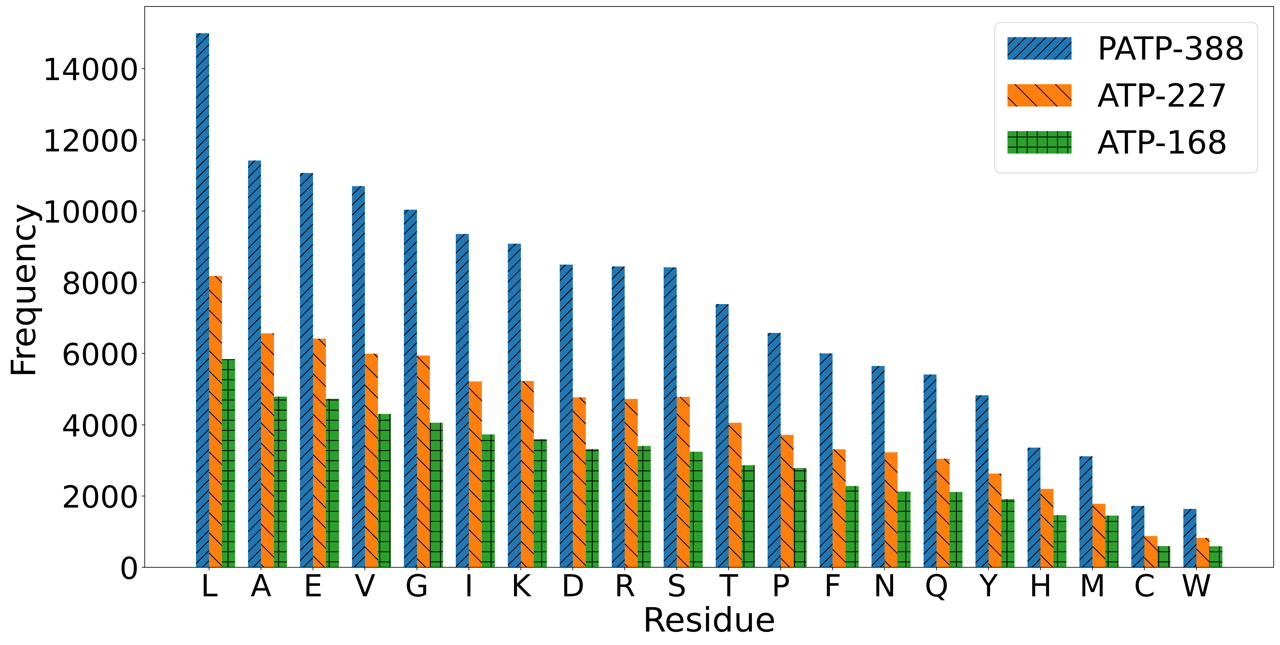}
        \caption{Frequency of ATP-binding residues in PATP-388, ATP-227 \& ATP-168}
        \label{fig:residue_freq}
\end{figure}

\begin{figure}[!htbp]
    \centering
        \includegraphics[width=1\linewidth]{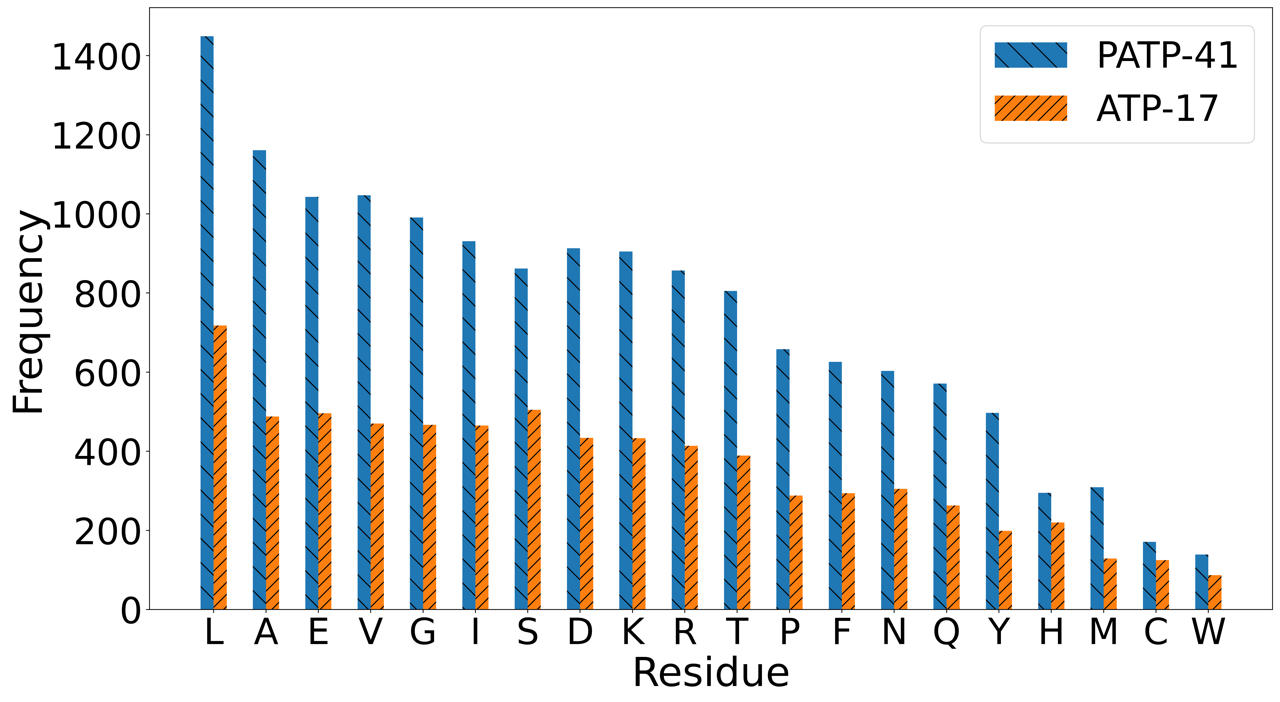}
        \caption{Frequency of ATP-binding residues in PATP-41(TEST) \& ATP-17}
        \label{fig:residue_freq_test}
\end{figure}

\begin{figure}[!htbp]
    \centering
    \includegraphics[width=1\linewidth]{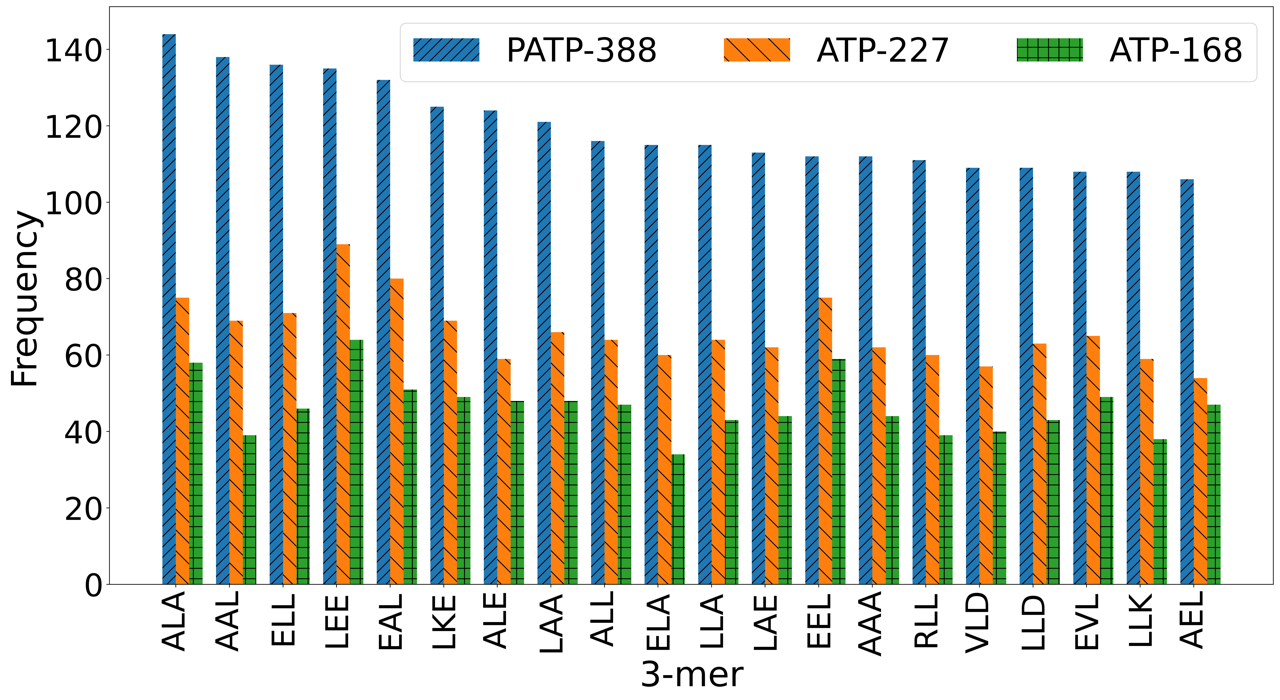}
    \caption{Top 20 ATP-binding residue 3-mers in PATP-388, ATP-227 \& ATP-168}
    \label{fig:3mer_freq}
\end{figure}

\begin{figure}[!htbp]
    \centering
    \includegraphics[width=1\linewidth]{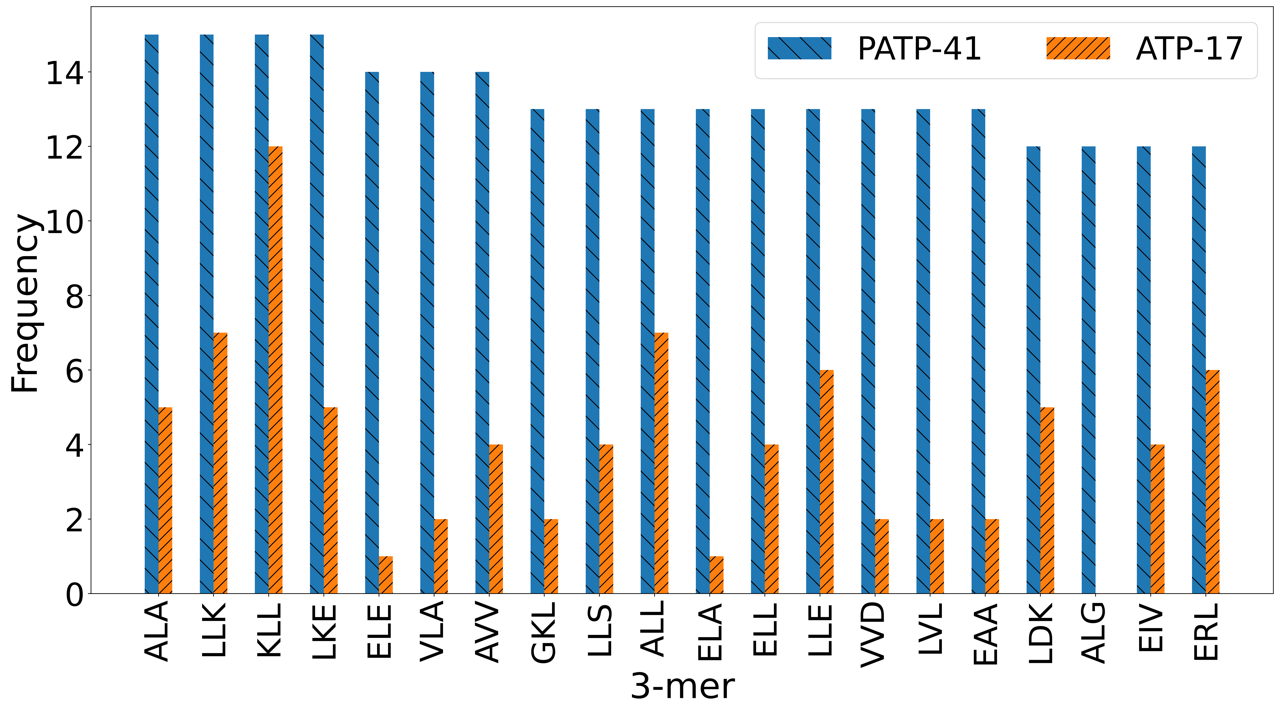}
    \caption{Top 20 ATP-binding residue 3-mers in PATP-41 \& ATP-17}
    \label{fig:3mer_freq_test}
\end{figure}

We observe some residues to be much more likely to be ATP-binding sites than others in experimental datasets. Figure \ref{fig:residue_freq} and \ref{fig:residue_freq_test} show the frequency plots of residues across three training and two test datasets, respectively. The same has been observed with the Test Datasets ATP-17 and PATP-41(TEST) also as seen in Figure \ref{fig:residue_freq_test}. The Leucine residues are seen to have a much higher affinity to bind to ATP molecules. A similar effect is observed at the 3-mer level (Figure \ref{fig:3mer_freq}, Figure \ref{fig:3mer_freq_test}). Many of the most frequent ATP-binding 3-mers have a Leucine residue in them. This high prevalence suggests a biochemical reason - the specific structural conformations adopted by these residues are favorable for interactions with ATP molecules. 
\section{Conclusion and future work}\label{sec:conclusion}
In this study, we propose a novel method for identifying ATP-binding sites that achieves comparable results to current predictors. The weights of the model occupy less than 30 MB, and the computation time for a single protein sequence is approximately 15 seconds. We have also analysed the ATP-binding residues and discovered that Leucine is frequently found at or near the binding sites. A suitable explanation, however, may be found through structural analysis. We are confident this work will be of use in tasks like protein function annotation and drug design. 
Utilising sequential modelling for the PSSM features by treating the protein as a sequence of PSSMs rather than a bag of unordered PSSMs, potentially with Conv-LSTM layers, could be one way to build upon this work. We plan to extend our work by exploring the utility of additional features, for example, solvent accessibility or using structural template-based methods such as TM-SITE. The work can be extended with the use of 3D structural data of the proteins. A further avenue for investigation involves the notion of ensembling the three models that have been evaluated in this study. The refinement of ATP binding predictions may be enhanced as a potential outcome.
Our technique can be scaled further by implementing more intricate neural network architectures and larger window sizes.
\bibliographystyle{aaai24}



\appendix
\section{Appendix}
\label{sec:appendix}
\subsection{Detailed Model Architecture}
\begin{figure}[!htbp]
    \centering
    \includegraphics[width=0.8\columnwidth]{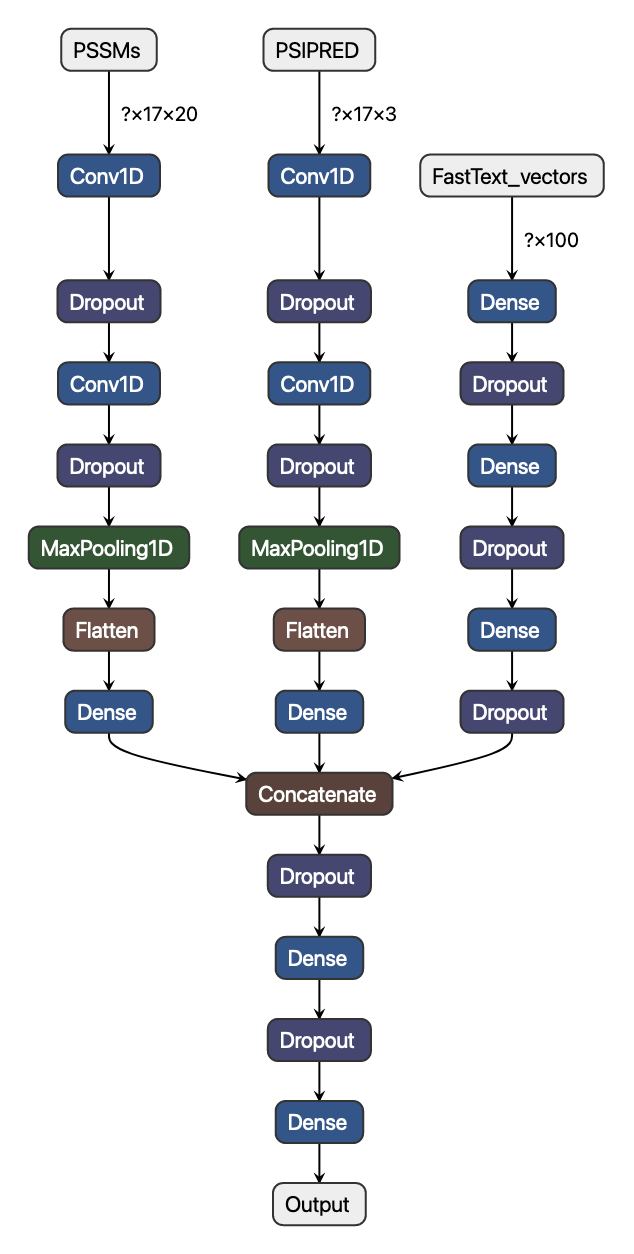}
    \caption{Detailed Model Architecture}
    \label{fig:detailed_model_architecture}
\end{figure}
\end{document}